\documentclass[11pt,epsf,letterpaper]{article}%
\usepackage{setspace}
\usepackage{amsmath}
\usepackage{color}
\usepackage{amsfonts}
\usepackage{verbatim}
\usepackage{amssymb}
\usepackage{graphicx}
\usepackage{amsmath}
\usepackage{amsfonts}
\usepackage{amssymb}
\usepackage{amsmath}%
\usepackage{amsmath}
\usepackage{amsfonts}
\usepackage{amssymb}

\providecommand{\U}[1]{\protect\rule{.1in}{.1in}}

\def\D{\tilde{\nabla}}

\title{Conformal couplings of a scalar field to higher curvature terms}
\author{Julio Oliva and Sourya Ray\\ \textit{Instituto de F\'{\i}sica, Facultad de Ciencias, Universidad
Austral de Chile, Valdivia, Chile.}\\{\small julio.oliva@docentes.uach.cl, ray@uach.cl}}

\begin{document}

\maketitle

\begin{abstract}
We present a simple way of constructing conformal couplings of a scalar field to higher order Euler densities. This is done by constructing a four-rank tensor involving the curvature and derivatives of the field, which transforms covariantly under local Weyl rescalings. The equation of motion for the field, as well as its energy momentum tensor are shown to be of second order. The field equations for the spherically symmetric ansatz are integrated, and for generic non-homogeneous couplings, the solution is given in terms of a polynomial equation, in close analogy with Lovelock theories.
\end{abstract}
\section{Introduction}

Conformally invariant theories have gained a lot of recent interest in various areas of physics and mathematics. One particular area of research is gravitational physics, where it is interesting to study the conformal coupling of a scalar field with gravity in arbitrary dimensions. Such a coupling was studied by Bocharova et al \cite{BBMB70} over forty years ago and later independently by Bekenstein \cite{Bekenstein:1974sf, Bekenstein:1975ts}, who found an exact black hole solution of Einstein equations in four spacetime dimensions. This solution is static, spherically symmetric, and asymptotically flat. Later this solution was generalized in arbitrary dimensions \cite{Xanthopoulos:1992fm} where it was shown that it represents a black hole only in four dimensions. In $D$ dimensions, the resulting equation of motion of the scalar field $\phi$ is given by
\begin{equation}
 \Box \phi - \frac{D-2}{4(D-1)}R \phi =0 \ .
\end{equation}
The second order operator on the left hand side, also known as the conformal Laplacian or the Yamabe operator, transforms covariantly under conformal transformations: $g_{ab}\rightarrow e^{2\Omega}g_{ab}$ and $\phi \rightarrow e^{1-\frac{D}{2}} \phi$ and plays an important role in the Yamabe problem \cite{Lee&Parker}.

It is natural to wonder if there are higher curvature generalizations of the conformal coupling of a scalar field to gravity. After a moment's thought one realizes that, for order $k \geq 2$, there is at least one trivial way of constructing such conformally invariant couplings of a scalar field. This can be done by simply taking $k$ conformal tensors and contracting all the indices with each other and then multiplying it by the scalar field $\phi$ raised to an appropriate power depending on the dimensions $D$. However, unlike the usual conformal coupling (linear in curvature), in this case the field equations are of fourth order. It turns out that there are other conformally invariant scalar densities, at each order $k$, out of which there is an unique density that leads to second order field equations (in dimensions $D > 2k$). These can be thought of as the {\it conformal couplings to Lovelock gravity} \cite{lovelock}. In this work, we give a simple way of constructing such scalar densities. We also obtain 
the corresponding energy momentum tensor and the general static spherically symmetric solutions of ${\cal T}^{(k)q}_p=0$ i.e., non-trivial solutions with vanishing energy-momentum tensor. This equation is conformally invariant and hence the solution is determined up to an arbitrary conformal factor. We shall show that these equations leave the scalar field undetermined. However, if one adds Lovelock terms as purely gravitational interactions in the action and looks for static spherically solutions of ${\cal G}^{(k)q}_p={\cal T}^{(k)q}_p=0$, then the scalar field is fixed and the general static spherically solution is given by the corresponding Lovelock solution. A similar construction was carried out in \cite{tekin}, in the case of quadratic theories, where an additional gauge field $A_\mu$ was introduced. Another construction was presented in \cite{Manvelyan:2006bk}, where however the resulting field equations are of higher order.

\section{Generalization of conformal coupling to Euler densities of arbitrary order}

In order to construct the conformal coupling of a scalar field to arbitrary higher powers of curvature, we follow the analogy with Lovelock gravity. We first construct a four rank tensor which is linear in the curvature and is conformally covariant  i.e., which transforms homogeneously under the following conformal transformations in arbitrary dimensions
\begin{equation}
 g_{ab}\rightarrow e^{2\Omega}g_{ab},\ \ \ \phi \rightarrow e^{s\Omega}\phi.
 \label{conftrans}
\end{equation}
One may see that the following tensor has the desired properties.
\begin{equation}
 S_{ij}^{kl}=\phi^2R_{ij}^{kl}+\frac{4}{s}\phi\delta_{[i}^{[k}\phi_{;j]}^{\ l]}+\frac{4(1-s)}{s^2}\delta_{[i}^{[k}\phi_{;j]}\phi_{;}^{\ l]}-\frac{2}{s^2}\delta_{[i}^{[k}\delta_{j]}^{l]}\phi_{;m}\phi_{;}^{\ m}\ .
 \label{confriem}
\end{equation}
where $\phi_{;i}=\nabla_{i}\phi$ and $\phi_{;j}^{\ k}=g^{ik}\nabla_i\nabla_j\phi$. Under the transformations (\ref{conftrans}) the above tensor transforms as
\begin{equation}
 S_{ij}^{kl} \rightarrow e^{2(s-1)\Omega} S_{ij}^{kl}\ .
\end{equation}
Now it is fairly easy to construct a conformally invariant coupling of a scalar field to arbitrary higher order Euler densities. For $s=1-D/2k$, the following density is conformally invariant
\begin{equation}
 I^{(k)}=\frac{1}{2^k}\int \sqrt{-g}d^{D}x\ \delta^{a_1b_1 \cdots a_kb_k}_{c_1d_1 \cdots c_kd_k}S^{c_1d_1}_{a_1b_1} \cdots S^{c_kd_k}_{a_kb_k}\ .
 \label{eulercoup}
\end{equation}
where $\delta^{a_1\cdots a_k}_{b_1\cdots b_k}=k!\delta^{[a_1}_{b_1}\cdots\delta^{a_k]}_{b_k}$ is the generalized Kronecker delta.
This action reduces to the usual conformally coupled scalar field for $k=1$. Note that the conformal weight $s$ is so chosen such that for each order $k$ there are no non-integral powers of $\phi$ in the Lagrangian in arbitrary dimensions. However, if one wishes to work with any other values of conformal weight then the Lagrangian has to be multiplied by appropriately compensating powers of the field $\phi$. In that case, one can consider the action
\begin{equation}
 I^{(k)}=\frac{1}{2^k}\int \sqrt{-g}d^{D}x\ \phi^{m_k} \delta^{a_1b_1 \cdots a_kb_k}_{c_1d_1 \cdots c_kd_k}S^{c_1d_1}_{a_1b_1} \cdots S^{c_kd_k}_{a_kb_k}\ .
 \label{eulercoup1}
\end{equation}
where $m_k=\dfrac{2k(1-s)-D}{s}$. Note that this action reduces to the $k$th order Lovelock action for a constant scalar field $\phi$. Varying the action with respect to the metric we obtain the following energy-momentum tensor.
\begin{align}
{\cal T}^{(k)q}_p=\frac{1}{2^{k+1}} \sqrt{-g}\ \phi^{m_k} \delta^{qa_1b_1 \cdots a_kb_k}_{pc_1d_1 \cdots c_kd_k}S^{c_1d_1}_{a_1b_1} \cdots S^{c_kd_k}_{a_kb_k}\ .
\end{align}
Whereas varying the scalar field $\phi$ we obtain the following equation of motion
\begin{align}
 \frac{m_k+2k}{2^{k}} \sqrt{-g}\ \phi^{m_k-1} \delta^{a_1b_1 \cdots a_kb_k}_{c_1d_1 \cdots c_kd_k}S^{c_1d_1}_{a_1b_1} \cdots S^{c_kd_k}_{a_kb_k}=0\ .
\end{align}
The equation of motion ensures that the trace of the energy-momentum tensor ${\cal T}^{(k)q}_p$ vanishes on its solutions as should be the case for conformally invariant theories. Also, note that all the field equations are of second order in the metric and the field (see Appendix B).

\section{Spherically symmetric solutions}

Let us now evaluate the field equation on a general spherically symmetric ansatz given by the following line element
\begin{align}
ds^{2}=\tilde{g}_{ij}(x)dx^{i}dx^{j}+e^{2\nu(x)}d\Sigma_{\gamma}^{2},%
\label{ansatz}
\end{align}
where $d\Sigma_{\gamma}^{2}=\hat{g}_{\alpha\beta}(y)dy^{\alpha}dy^{\beta}$ is
the line element of a $(D-2)$-dimensional space of constant curvature $\gamma
$. Let $\D$ be the Levi-Civita connection on the two-dimensional space orthogonal to the constant curvature space and $\tilde{R}$ be the corresponding scalar curvature. To evaluate the field equations it is more convenient to express the scalar field as
\begin{equation}
 \phi=e^{-s\lambda}.
\end{equation}
Then the energy-momentum tensor and the equation of motion can be respectively written as
\begin{align}
{\cal T}^{(k)q}_p=\frac{1}{2^{k+1}} \sqrt{-g}\ e^{(D-2k)\lambda} \delta^{qa_1b_1 \cdots a_kb_k}_{pc_1d_1 \cdots c_kd_k}Z^{c_1d_1}_{a_1b_1} \cdots Z^{c_kd_k}_{a_kb_k}\ .
\end{align}
and
\begin{align}
 \frac{m_k+2k}{2^{k}} \sqrt{-g}\ e^{(D-2k+s)\lambda} \delta^{a_1b_1 \cdots a_kb_k}_{c_1d_1 \cdots c_kd_k}Z^{c_1d_1}_{a_1b_1} \cdots Z^{c_kd_k}_{a_kb_k}=0\ .
\end{align}
where
\begin{equation}
 Z_{ab}^{cd}=R_{ab}^{cd}-4\delta_{[a}^{[c}\lambda_{;b]}^{\ d]}+4\delta_{[a}^{[c}\lambda_{;b]}\lambda_{;}^{\ d]}-2\delta_{[a}^{[c}\delta_{b]}^{d]}\lambda_{;m}\lambda_{;}^{\ m}\ .
\end{equation}
Then the nontrivial components of the Riemann curvature tensor and the tensor $Z_{ab}^{cd}$ are given by
\begin{align}
& R_{jl}^{\ \ ik}=\frac{1}{2}\tilde{R}\delta_{jl}^{ik},\qquad \qquad \qquad \qquad Z_{jl}^{\ \ ik}=\frac{1}{2}(\tilde{R}-2(\D_{m}\lambda)(\D^{m}\lambda))\delta_{jl}^{ik}-4\delta_{[j}^{[i}\tilde{\mathcal{C}^{\prime}}^{k]}_{l]},&\\
& R_{\nu\rho}^{\ \ \mu\lambda}=\tilde{\mathcal{B}}\delta_{\nu\rho}^{\mu\lambda},\qquad \qquad \qquad \qquad Z_{\nu\rho}^{\ \ \mu\lambda}=\tilde{\mathcal{B}^{\prime}}\delta_{\nu\rho}^{\mu\lambda},&\\
& R_{j\nu}^{\ \ i\mu}=-\tilde{\mathcal{A}}_{j}^{i}\delta_{\nu
}^{\mu} \qquad \qquad \qquad \qquad Z_{j\nu}^{\ \ i\mu}=-\tilde{\mathcal{A}^{\prime}}_{j}^{i}\delta_{\nu
}^{\mu}&.
% &  C_{jl}^{\ \ ik}=\frac{(D-3)\tilde{S}}{2(D-1)}\delta_{jl}^{ik},\qquad C_{\nu\rho
% }^{\ \ \mu\lambda}=\frac{\tilde{S}}{(D-1)(D-2)}\delta_{\nu\rho}^{\mu\lambda},\qquad
% C_{j\nu}^{\ \ i\mu}=-\frac{(D-3)\tilde{S}}{2(D-1)(D-2)}\delta_{j}^{i}\delta_{\nu
% }^{\mu},&
\end{align}
where
\begin{align*}
&\tilde{\mathcal{B}}=\gamma e^{-2\nu}-(\D_{m}\nu
)(\D^{m}\nu), \qquad \qquad \qquad \tilde{\mathcal{B}^\prime}=\tilde{\mathcal{B}}-2(\D_k\nu)(\D^k\lambda)-(\D_k\lambda)(\D^k\lambda)&\\
&\tilde{\mathcal{A}}_{j}^{i}=\D^{i}%
\D_{j}\nu+(\D^{i}\nu)(\D_{j}\nu), \qquad \qquad \qquad \tilde{\mathcal{A}^\prime}_{j}^{i}=\tilde{\mathcal{A}}_{j}^{i}+\tilde{\mathcal{C}^{\prime}}_{j}^{i}+\delta_{j}^{i}\D^k\lambda \D_k(\nu+\lambda)&\\
&\ \qquad \qquad \qquad \qquad  \qquad \qquad  \qquad \quad \text{and}\ \ \ \tilde{\mathcal{C}^{\prime}}_{j}^{i}=\D_j\D^i\lambda-\D_j\lambda \D^i\lambda .&
\end{align*}
Then the non-vanishing components of the energy-momentum tensor evaluated on the ansatz (\ref{ansatz}) are
\begin{align}
& \mathcal{T}^{(k)i}_j=\dfrac{\sqrt{-g}\ e^{(D-2k)\lambda}(D-2)!}{2(D-2k-1)!}\tilde{\mathcal{B}^\prime}^{k-1}\left[(D-2k-1)\tilde{\mathcal{B}^\prime}\delta^i_j-2k\delta^{ik}_{jl}\tilde{\mathcal{A}^\prime}^l_k\right] \label{em1}\\
&\mathcal{T}^{(k)\alpha}_{\beta}=\dfrac{\sqrt{-g}\ e^{(D-2k)\lambda}(D-3)!}{2(D-2k-1)!}\tilde{\mathcal{B}^\prime}^{k-2}\delta^{\alpha}_{\beta}\left[(D-2k-1)(D-2k-2)\tilde{\mathcal{B}^\prime}^2+k\tilde{\mathcal{B}^\prime}(\tilde{R}-2\D_{m}\D^{m}\lambda  \right. \nonumber \\ & \qquad-2(D-2k-1)\tilde{\mathcal{A}^\prime}^i_i)
 \left.+2k(k-1)\delta^{ik}_{jl}\tilde{\mathcal{A}^\prime}^j_i\tilde{\mathcal{A}^\prime}^l_k\right] \label{em2}
\end{align}
and the equation of motion or the trace of the energy momentum tensor is given by
\begin{eqnarray}
& \dfrac{\sqrt{-g}\ e^{(D-2k+s)\lambda}(D-2)!}{(D-2k-1)!s}\tilde{\mathcal{B}^\prime}^{k-2}\left[(D-2k)(D-2k-1)\tilde{\mathcal{B}^\prime}^2+k\tilde{\mathcal{B}^\prime}(\tilde{R}-2\D_{m}\D^{m}\lambda-2(D-2k)\tilde{\mathcal{A}^\prime}^i_i)\right. \nonumber \\
& \qquad \left. +2k(k-1)\delta^{ik}_{jl}\tilde{\mathcal{A}^\prime}^j_i\tilde{\mathcal{A}^\prime}^l_k\right]=0 \label{eom}
\end{eqnarray}
Obviously, for $k > 2$ there is trivial solution to the equations (\ref{em1}), (\ref{em2}) and (\ref{eom}) which is $\tilde{\mathcal{B}^\prime}=0$ i.e.,
\begin{equation}
 \gamma e^{-2\nu}-\D_{m}(\nu+\lambda)\D^{m}(\nu+\lambda)=0.
\label{degeq}
\end{equation}
Note that since the field equations are invariant with respect to local Weyl rescalings one can as well gauge away the warp factor in front of the constant curvature base manifold in (\ref{ansatz}) by a conformal transformation. This implies that we can choose $\nu$ to be zero without any loss of generality. In that case, the previous equation becomes
$(\D_k\lambda)(\D^k\lambda)=\gamma$. This is the Hamilton-Jacobi equation in curved spacetime. For $\gamma=0$, the general solution of the equation (\ref{degeq}) is $\lambda=f(u)$ or $\lambda=g(v)$, where $(u,v)$ are the null coordinates in the two dimensional space. For $\gamma \neq 0$, there is no general solution known. We next analyze the non-trivial case $\tilde{\mathcal{B}^\prime}\neq 0$. We first show that the field equations imply that the metric is static. To see this, contract the index $i$ after multiplying the equation obtained from (\ref{em1}) by $\epsilon_{ki}$ and then symmetrize the indices $(j,k)$ to obtain
\begin{align}
 e^{\lambda}\D_{(j}\left[\epsilon_{k)i}\D^ie^{-\lambda}\right]=0
\end{align}
In other words, the vector $\xi_k=\epsilon_{ki}\D^ie^{-\lambda}$ is a Killing vector.
\begin{itemize}
 \item {\bf Case I}: $\xi^k$ is null i.e., $(\D^k \lambda)(\D_k \lambda)=0$
\end{itemize}
In this case we also assume that $\gamma \neq 0$ since otherwise we are led back to the trivial case of $\tilde{\mathcal{B}^\prime}=0$. Now, taking the trace of $\mathcal{T}^{(k)i}_j=0$ and plugging it back we obtain $\tilde{\mathcal{A}^\prime}^k_k\delta^{i}_{j}=2\tilde{\mathcal{A}^\prime}^i_j$. If $\xi^k$ is null then this implies $D=2k+1$. Furthermore, the $(\alpha,\beta)$-components of the field equations imply that the two dimensional metric $\tilde{g}_{ij}$ is flat and $e^{-\lambda}=C_1u+C_2$ or $C_1v+C_2$ where $(u,v)$ are the light-cone coordinates and $C_1$ and $C_2$ are arbitrary constants.
\begin{itemize}
 \item {\bf Case II}: $\xi^k$ is non-null i.e., $(\D^k \lambda)(\D_k \lambda)\neq0$
\end{itemize}
In this case we introduce one of the coordinates $R=e^{-\lambda}$ and the other $t$ such that $\xi^k=\left(\dfrac{\partial}{\partial t}\right)^k$. We then use the following metric ansatz
\begin{equation}
 ds^{2}=-f(R)dt^2+\dfrac{dR^2}{g(R)}+d\Sigma_{\gamma}^{2},%
 \label{ansatz1}
\end{equation}
In this case the $(t,R)$ and $(R,t)$ components of the field equations (\ref{em1}) are trivially satisfied. The equations $\mathcal{T}^{(k)t}_t-\mathcal{T}^{(k)R}_R=0$ imply $f(R)=\kappa g(R)$, where $\kappa$ is a constant and can be absorbed by redefining the time coordinate. Finally the equation $\mathcal{T}^{(k)t}_t+\mathcal{T}^{(k)R}_R=0$ can then be expressed as
\begin{align}
 (D-2k-1)\tilde{\mathcal{B}^\prime}-kR\dfrac{d\tilde{\mathcal{B}^\prime}}{dR}=0 \qquad \text{where}\ \ \tilde{\mathcal{B}^\prime}=\left(\gamma-\dfrac{g(R)}{R^2}\right),
\end{align}
which can be integrated to obtain
\begin{align}
 f(R)=g(R)=\gamma R^2-CR^{\frac{D-1}{k}}
\end{align}
and the scalar field is then given by $\phi=R^s$. This solves the equation (\ref{em2}) trivially. The metric can as well be given in terms of the scalar field $\phi$ in the following way
\begin{align}
 ds^{2}=-f(\phi)dt^2+\dfrac{d\phi^2}{g(\phi)}+d\Sigma_{\gamma}^{2},%
\end{align}
where
\begin{eqnarray*}
 &f(\phi)=\phi^{2/s}\left(\gamma-C\phi^{\frac{D-2k-1}{sk}}\right)\\
 &g(\phi)=s^2\phi^2\left(\gamma-C\phi^{\frac{D-2k-1}{sk}}\right)
\end{eqnarray*}
The scalar field $\phi$ is then left completely arbitrary.

\section{Non-homogeneous couplings}

Now it is fairly easy to consider non-homogeneous conformal couplings of a scalar field to higher order Lovelock densities. Consider the action
\begin{equation}
 I=\sum_kc_kI^{(k)}=\sum_k\frac{1}{2^k}\int \sqrt{-g}d^{D}x\ c_k\phi^{m_k} \delta^{a_1b_1 \cdots a_kb_k}_{c_1d_1 \cdots c_kd_k}S^{c_1d_1}_{a_1b_1} \cdots S^{c_kd_k}_{a_kb_k}\ .
\end{equation}
in which case the field equations and the equation of motion of the scalar field are given by
\begin{align}
&{\cal T}^{q}_p=\sum_kc_k{\cal T}^{(k)q}_p=\sum_k\frac{c_k}{2^{k+1}} \sqrt{-g}\ \phi^{m_k} \delta^{qa_1b_1 \cdots a_kb_k}_{pc_1d_1 \cdots c_kd_k}S^{c_1d_1}_{a_1b_1} \cdots S^{c_kd_k}_{a_kb_k}\\
&\sum_kc_k \frac{m_k+2k}{2^{k}} \sqrt{-g}\ \phi^{m_k-1} \delta^{a_1b_1 \cdots a_kb_k}_{c_1d_1 \cdots c_kd_k}S^{c_1d_1}_{a_1b_1} \cdots S^{c_kd_k}_{a_kb_k}=0\ .
\end{align}
Evaluating these equations on the spherically symmetric ansatz (\ref{ansatz}), we obtain
\begin{align}
& \mathcal{T}^{i}_j=\sum_k\dfrac{\sqrt{-g}\ e^{(D-2k)\lambda}\hat{c}_k}{2(D-1)}\tilde{\mathcal{B}^\prime}^{k-1}\left[(D-2k-1)\tilde{\mathcal{B}^\prime}\delta^i_j-2k\delta^{ik}_{jl}\tilde{\mathcal{A}^\prime}^l_k\right] \label{em10}\\
&\mathcal{T}^{\alpha}_{\beta}=\sum_k\dfrac{\sqrt{-g}\ e^{(D-2k)\lambda}\hat{c}_k}{2(D-1)(D-2)}\tilde{\mathcal{B}^\prime}^{k-2}\delta^{\alpha}_{\beta}\left[(D-2k-1)(D-2k-2)\tilde{\mathcal{B}^\prime}^2+k\tilde{\mathcal{B}^\prime}(\tilde{R}-2\D_{m}\D^{m}\lambda  \right. \nonumber \\ & \qquad-2(D-2k-1)\tilde{\mathcal{A}^\prime}^i_i)
 \left.+2k(k-1)\delta^{ik}_{jl}\tilde{\mathcal{A}^\prime}^j_i\tilde{\mathcal{A}^\prime}^l_k\right] \quad \text{where}\ \hat{c}_k=\dfrac{(D-1)!}{(D-2k-1)!}c_k\label{em11}
\end{align}
whereas the equation of motion take the form
\begin{align}
& \sum_k\dfrac{\sqrt{-g}\ e^{(D-2k+s)\lambda}\hat{c}_k}{(D-1)s}\tilde{\mathcal{B}^\prime}^{k-2}\left[(D-2k)(D-2k-1)\tilde{\mathcal{B}^\prime}^2+k\tilde{\mathcal{B}^\prime}(\tilde{R}-2\D_{m}\D^{m}\lambda-2(D-2k)\tilde{\mathcal{A}^\prime}^i_i)\right. \nonumber \\
& \qquad \left. +2k(k-1)\delta^{ik}_{jl}\tilde{\mathcal{A}^\prime}^j_i\tilde{\mathcal{A}^\prime}^l_k\right]=0 \label{eom1}
\end{align}
Now, again using the conformal invariance of the system we fix the gauge $\nu=0$ and proceeding as before we can show that $\xi_k=\epsilon_{ki}\D^ie^{-\lambda}$ is a Killing vector, provided $\sum_kk\hat{c}_k(\tilde{\mathcal{B}^\prime} e^{-2\lambda})^{k-1}\neq0$.
\begin{itemize}
 \item {\bf Case I}: $\xi^k$ is null i.e., $(\D^k \lambda)(\D_k \lambda)=0$
\end{itemize}
Again assuming $\gamma \neq0$, in this case the remaining field equations imply both the field $\lambda$ and $\tilde{R}$ are constants.
\begin{itemize}
 \item {\bf Case II}: $\xi^k$ is non-null i.e., $(\D^k \lambda)(\D_k \lambda)\neq0$
\end{itemize}
Introducing the coordinates $R=e^{-\lambda}$ and $t$ such that $\xi^k=\left(\dfrac{\partial}{\partial t}\right)^k$, we evaluate the field equations on the ansatz (\ref{ansatz1}). The $(t,R)$ and $(R,t)$ components of the field equations (\ref{em10}) are then trivially satisfied. Furthermore, the equation $\mathcal{T}^{t}_t-\mathcal{T}^{R}_R=0$ imply $f(R)=\kappa g(R)$, where $\kappa$ is a constant and can be absorbed by redefining the time coordinate. Finally the equation $\mathcal{T}^{(k)t}_t+\mathcal{T}^{(k)R}_R=0$ can then be expressed as
\begin{align}
 \sum_k\hat{c}_kR^{-(D-2k)}\left[(D-2k-1)\tilde{\mathcal{B}^\prime}^k-kR\tilde{\mathcal{B}^\prime}^{k-1}\dfrac{d\tilde{\mathcal{B}^\prime}}{dR}\right]=0 \qquad \text{where}\ \ \tilde{\mathcal{B}^\prime}=\left(\gamma-\dfrac{g(R)}{R^2}\right) \label{em13}
\end{align}
which can be rewritten as
\begin{align}
\dfrac{d}{dR}\sum_k\hat{c}_k\{R^{-(D-1)}(\tilde{\mathcal{B}^\prime}R^{2})^k\}=0
\end{align}
which can then be integrated as
\begin{align}
\sum_k\hat{c}_k(\tilde{\mathcal{B}^\prime}R^{2})^k=C_1R^{(D-1)}
\label{poly}
\end{align}

% Then as previously, assuming $\tilde{\mathcal{B}^\prime}\neq 0$ and $\lambda=\lambda(r)$, the $(t,r)$ and $(r,t)$ components of the field equations (\ref{em10}) implies that $g(r,t)=g(r)$ provided $\sum_kk\hat{c}_k\tilde{\mathcal{B}^\prime}^{k-1}\neq0$. Consequently the equations $\mathcal{T}^{t}_t-\mathcal{T}^{r}_r=0$ and $\mathcal{T}^{t}_t+\mathcal{T}^{r}_r=0$ respectively can be expressed as
% \begin{align}
% &(\gamma-\tilde{\mathcal{B}^\prime})\left(1+\dfrac{\dot{f}}{2f\dot{\lambda}}\right)=-\dfrac{1}{2}\dfrac{d\tilde{\mathcal{B}^\prime}}{d\lambda} \label{em12}\\
% &\sum_k\hat{c}_ke^{(D-2k)\lambda}\left[(D-2k-1)\tilde{\mathcal{B}^\prime}^k+k\tilde{\mathcal{B}^\prime}^{k-1}\dfrac{d\tilde{\mathcal{B}^\prime}}{d\lambda}\right]=0 \qquad \text{where}\ \ \tilde{\mathcal{B}^\prime}=(\gamma-g(r)(\dot{\lambda}(r))^2) \label{em13}
% \end{align}
% where we have assumed, without any loss of generality $f=f(r)$. Now, one can integrate the equation (\ref{em13}) by rewriting it as
% \begin{align}
% e^{\lambda}\dfrac{d}{d\lambda}\sum_k\hat{c}_k\{e^{(D-1)\lambda}(\tilde{\mathcal{B}^\prime}e^{-2\lambda})^k\}=0
% \end{align}
% which can then be integrated as
% \begin{align}
% \sum_k\hat{c}_k(\tilde{\mathcal{B}^\prime}e^{-2\lambda})^k=C_1e^{-(D-1)\lambda}
% \label{poly}
% \end{align}
This is a polynomial equation in $\tilde{\mathcal{B}^\prime}$. We may solve this polynomial to obtain $\tilde{\mathcal{B}^\prime}=\tilde{\mathcal{B}^\prime}(R)$, which in turn allows one to express the metric functions as
\begin{align}
&g(R)=f(R)=R^2\left(\gamma-\tilde{\mathcal{B}^\prime}(R)\right)
\end{align}
This satisfies the $(\alpha,\beta)$ components of the field equations. Again note that the scalar field is then given by $\phi=R^s$. As before, one can write the metric in terms of the scalar field $\phi$ in the following way
\begin{align}
 ds^{2}=-f(\phi)dt^2+\dfrac{d\phi^2}{g(\phi)}+d\Sigma_{\gamma}^{2},%
\end{align}
where
\begin{eqnarray*}
 &f(\phi)=\phi^{2/s}\left(\gamma-C\tilde{\mathcal{B}^\prime}(\phi^{1/s})\right)\\
 &g(\phi)=s^2\phi^2\left(\gamma-C\tilde{\mathcal{B}^\prime}(\phi^{1/s})\right)
\end{eqnarray*}
The scalar field $\phi$ is then left completely arbitrary.

However, when one adds a purely gravitational interaction in the action which is not conformally invariant then the field equations are no longer conformally invariant. In this case, one cannot gauge away the warp factor in front of the $(D-2)$ dimensional constant curvature base manifold. Moreover, the scalar field $\phi$ is then determined by the field equations.

\section{Conclusions}
 Here we have presented a novel construction of conformal couplings of a scalar field to arbitrary higher order Euler densities. This is done by first constructing a four rank tensor linear in the curvature which transforms covariantly under conformal transformations and has the symmetries of the Riemann tensor (except the Bianchi identity). This tensor along with the generalized Kronecker delta is then used to construct conformal invariants of higher order in parallel with the construction of Euler densities. The resulting energy momentum tensor is shown to be of second order. We further solve the equations of motion under spherically symmetric conditions.

 Let us now briefly mention some of the potential future directions of study where the present work could be of some relevance.

 Firstly, as mentioned in the introduction, the usual conformally coupled scalar field was originally studied in the context of black hole no-hair theorems. The BBMB or Bekenstein black hole circumvents the no-hair theorem since the scalar field diverges at the horizon. However, it was shown that the higher dimensional generalization of this solution does not represent a black hole \cite{Xanthopoulos:1992fm}. Thus the Bekenstein black hole is the only known, asymptotically flat, static black hole with a conformal scalar hair \footnote{Recently a spacetime belonging to the family of the Plebanski-Demianski metrics was found to be a solution of the system \cite{anabalonmaeda}, which reduces to the C-metric in the nonrotating case (also found in \cite{leftcharm}). It is interesting to note, that the usual conical singularity in the C-metric is removed by the presence of the scalar field.}. So, it is natural to look for black hole with a conformal scalar hair in higher dimensions where the coupling involve 
higher curvature terms.

 Secondly, the scalar fields discussed here falls into the class of Galileons which are scalars whose equations of motion depend only on second derivatives. Hence in flat space these are invariant under constant shifts of the fields and their gradients. These field have gain some attention in the community due to various intriguing properties and their applications in particle physics and cosmology (see e.g. \cite{galileonsi}-\cite{galileonsf}). It has also been shown that the Galileons can be obtained through a standard Kaluza-Klein reduction of higher order Lovelock gravity \cite{galileonsfromkklov}. This naturally raises a question about the compactification such that the Galileons have a conformal invariance.

 Thirdly, one may notice that there is a one-to-one correspondence of the spherically symmetric solutions of Lovelock theories to those of ${\cal T}^{(k)q}_p=0$. However, in general Lovelock theories in odd dimensions, there is an enhancement of symmetry when the coupling constants are tuned such that the theory has a unique vacuum. This happens when the corresponding polynomial satisfied by the unknown metric function in static coordinates has a unique solution. Specifically, in odd dimensions, it has been shown that the theory can then be written as a Chern-Simons gauge theory \cite{cham}, \cite{zanelli}. So, it is natural to investigate the role of any such enhancement of symmetry when the polynomial (\ref{poly}) has a unique solution.

 Finally, even though in this work we have considered conformal couplings to Euler densities only, one can also use the tensor (\ref{confriem}) to construct other interesting couplings.

 \section*{Acknoledgements}
We thank Fabrizio Canfora, Christos Charmousis, Mokhtar Hassaine and Jorge Zanelli for enlightening comments. This work is partially
supported by FONDECYT grants 11090281, and 11110176, and by CONICYT grant 791100027.

\appendix

\section{Examples}

Here we provide the explicit expressions for the conformal couplings to the Euler densities up to the first few orders. Let us choose the conformal weight to be $s=1-D/2$. Then, the coupling to the cosmological constant is given by
\begin{equation}
 \sqrt{-g}\ \phi^{\frac{2D}{D-2}}\ ,
\end{equation}
that to the Einstein-Hilbert term is
\begin{equation}
 \frac{1}{2}\sqrt{-g}\ \delta^{cd}_{ab}S^{ab}_{cd}=\sqrt{-g}\left[R\phi^2-\frac{4(D-1)}{D-2}\phi\Box\phi\right]\ ,
\end{equation}
that to the Gauss-Bonnet term is
\begin{eqnarray}
&& \frac{1}{4}\sqrt{-g}\ \phi^{-\frac{2D}{D-2}}\delta^{c_1d_1c_2d_2}_{a_1b_1a_2b_2}S^{a_1b_1}_{c_1d_1}S^{a_2b_2}_{c_2d_2}\nonumber\\
&&= \sqrt{-g}\ \phi^{-\frac{2D}{D-2}}\biggl[\left(R_{ab}^{\ \ cd}R_{cd}^{\ \ ab}-4R_{ab}R^{ab}+R^2\right)\phi^4 \biggr. \nonumber\\
&&+8\frac{D-3}{(D-2)}\left(2\phi^2 R_{ab}(\phi\nabla^a\nabla^b\phi-\frac{D}{D-2}\nabla^a\phi\nabla^b\phi) \right. \nonumber\\
&& -\phi^2 R(\phi \Box \phi-\frac{2}{D-2}\nabla_a\phi\nabla^a\phi)-2\phi^2\nabla_a\nabla_b\phi(\nabla^a\nabla^b\phi-g^{ab}\Box \phi)\nonumber\\
&&\left. \left.+\frac{4}{D-2}\phi\nabla_a\phi(D\nabla_b\phi\nabla^b\nabla^a\phi-\nabla^a\phi\Box\phi)-\frac{2D(D-1)}{(D-2)^2}(\nabla_a\phi\nabla^a\phi)^2\right)\right]\ .
\label{gbcoupling}
\end{eqnarray}

\section{A useful identity}

As noted previously, the energy momentum tensor for the scalar field is second order in the fields. This can be checked by using the following identity. First let us define a four rank tensor
\begin{eqnarray}
 X^{a_1b_1}_{c_1d_1}= \delta^{a_1b_1 \cdots a_kb_k}_{c_1d_1 \cdots c_kd_k}Z^{c_2d_2}_{a_2b_2} \cdots Z^{c_kd_k}_{a_kb_k}\ .
\end{eqnarray}
Then 
\begin{eqnarray}
 \nabla_aX^{ab}_{cd}=-(D-2k+1)X^{eb}_{cd}\lambda_{;e}+2\delta^e_a X^{ab}_{e[d}\lambda_{;c]} \ .
\end{eqnarray}

\end{document}